# Inhomogeneous Kondo-lattice in geometrically frustrated $Pr_2Ir_2O_7$

Mariam Kavai[1]†, Joel Friedman[1]†, Kyle Sherman[1]†, Mingda Gong[1], Ioannis Giannakis[1], Samad Hajinazar[1], Haoyu Hu[2], Sarah E. Grefe[2], Justin Leshen[1], Qiu Yang[3], Satoru Nakatsuji[3-5], Aleksey N. Kolmogorov[1], Qimiao Si[2], Michael Lawler[1], and Pegor Aynajian[1]*

[1]Department of Physics, Applied Physics and Astronomy, Binghamton University, Binghamton, New York 13902, USA
[2]Department of Physics and Astronomy, Rice Center for Quantum Materials, Rice University, Houston, Texas, 77005, USA
[3]Institute for Solid State Physics, University of Tokyo, Kashiwa, Chiba 277-8581, Japan
[4] Department of Physics, University of Tokyo, Bunkyo-ku, Tokyo 113-0033, Japan
[5]Trans-scale Quantum Science Institute, University of Tokyo, Bunkyo-ku, Tokyo 113-0033, Japan

* To whom correspondence should be addressed: aynajian@binghamton.edu
†Equal contribution

**Magnetic fluctuations induced by geometric frustration of local Ir-spins disturb the formation of long range magnetic order in the family of pyrochlore iridates, $R_2Ir_2O_7$ (R = lanthanide) [1]. As a consequence, $Pr_2Ir_2O_7$ lies at a tuning-free antiferromagnetic-to- paramagnetic quantum critical point and exhibits a diverse array of complex phenomena including Kondo effect, biquadratic band structure, metallic spin-liquid (MSL), and anomalous Hall effect[2–5]. Using spectroscopic imaging with the scanning tunneling microscope, complemented with machine learning K-means clustering analysis, density functional theory, and theoretical modeling, we probe the local electronic states in single crystal of $Pr_2Ir_2O_7$ and discover an electronic phase separation. Nanoscale regions with a well-defined Kondo resonance are interweaved with a non-magnetic metallic phase with Kondo-destruction . Remarkably, the spatial nanoscale patterns display a correlation-driven fractal geometry with power-law behavior extended over two and a half decades, consistent with being in proximity to a critical point. Our discovery reveals a new nanoscale tuning route, viz. using a spatial variation of the electronic potential as a means of adjusting the balance between Kondo entanglement and geometric frustration.**

A Kondo lattice, a matrix of localized spins embedded in a sea of itinerant electrons, is traditionally understood as giving rise to maximally-entangled singlets between the two types of spins, leading to a Heavy Fermi liquid (HFL). When it is exposed to strong geometric frustration, the enhanced quantum fluctuations among the local moments are expected to break down the Kondo effect[6–8] and may lead to the emergence of a novel and rare phase of quantum matter[8–14] - a MSL[3]. Analogous to the elusive quantum spin-liquid anticipated in frustrated magnetic insulators, this exotic phase of quantum matter may host unusual fractionalized excitations and bare the essence of unconventional superconductivity[15,16].

Over the past decade, theoretical efforts focused on the description of a unified global magnetic phase diagram that merges geometric frustration (G) and Kondo hybridization ($J_K$)[9–11] (see Fig.1a). In the limit of small G and $J_K$, local moments form a long-range magnetic order governed by the Ruderman–Kittel–Kasuya–Yosida (RKKY) interaction. However, the magnetic order becomes unstable as either G or $J_K$ is enhanced. The regime of large $J_K$ has been sufficiently investigated in the past, both theoretically and experimentally[6,7,17–19]. Simply put, the Kondo entanglement delocalizes the $f$-electrons and the resulting HFL has a large Fermi surface that incorporates both the itinerant- and $f$-electrons.

The regime of strong geometric frustration (large G) is not yet well established. In this regime, long-range antiferromagnetic order is no longer sustained due to enhanced quantum fluctuations, giving way to a MSL ground state that preserves the long-range entanglement of spin singlets. In this phase, the spins are localized and the Fermi surface is determined by the itinerant conduction electrons only, characterized by the small Fermi surface[6–8]. As the HFL and MSL involve the same local magnetic moments and therefore locally compete, the mutual territory of strong geometric frustration and Kondo hybridization (large G and large $J_K$) may give rise to unconventional forms of quantum criticality. Experimentally, this territory is largely unexplored[12–14].

From a materials perspective, the pyrochlore iridate $Pr_2Ir_2O_7$ is one of the very few existing systems that may enable us to investigate this regime[3,20]. The $4f^2$-electron moments of the $Pr^{3+}$ are located on the edges of the corner-sharing tetrahedra of the pyrochlore structure, forming a Kagome lattice of Ising spins pointing along the [111] direction[21]. Susceptibility measurements show the $4f$-moments to have an AFM RKKY interaction of T* ~ 20 K mediated by Ir 5d-conduction electrons[3]. Yet, $Pr_2Ir_2O_7$ is the only member of the pyrochlore iridate family that remains magnetically disordered and metallic down to the lowest measured temperature[21,22] and is argued to be located in proximity to a quantum critical point[2,23]. The 5$d$ electrons of the $Ir^{4+}$ states, which are in the $J_{eff}$ = ½ state, are semi-metallic and form a quadratic band-touching[24] at the zone center in $Pr_2Ir_2O_7$ as confirmed by angle resolved photoemission spectroscopy (ARPES)[4]. This is in contrast to most other members of the pyrochlore family, where they lead to an insulating AFM [3,21–23]. These metallic states, therefore, mediate Kondo hybridization of the Pr 4$f$-moments[25]. The non-Kramers doublet ground state of the J = 4 Pr orbitals indicates a possible two-channel Kondo effect with the excited Kramers doublet[21]. Indeed, resistivity measurements show a minimum and a weak upturn below 40 K providing evidence of Kondo behavior[3]. On the other hand, the magnetic susceptibility starts to diverge below about 40 K[3] suggesting that $4f$ moments, at least partially, are not Kondo-quenched, rather fluctuating down to the lowest temperature as a result of the magnetic frustration. Below T = 1.7 K, the susceptibility follows a Ln*T* dependence indicating a spin-liquid behavior[3]. Partial spin freezing is observed only below $T_f$ = 120 mK[3]. This makes $Pr_2Ir_2O_7$ a distinct example of a magnetically frustrated Kondo-lattice that may display MSL behavior.

Here we use spectroscopic imaging with the scanning tunneling microscope to spatially visualize and probe the local electronic states in single crystal of $Pr_2Ir_2O_7$[26]. To our knowledge, this is the only STM experiment on any pyrochlore iridate system to date. Figure 1b shows a schematic of the crystal structure revealing the Kagome lattice of Pr- and Ir-rich layers alternating along the [111]-direction[27]. The lack of an easy cleaving plane in this cubic material system with strong interlayer metal-oxygen bonds, as examined in our density functional theory (DFT) calculations, makes STM experiments particularly challenging. Over 20 different attempts of single crystal cleaves have been carried out to achieve the desired atomically resolved surface topographs. Fig.1c shows a topographic STM image (65 nm) of the cleaved (111) surface revealing the atomically resolved lattice (see Supplementary Note 1 for more detail) with an in-plane spacing of 7.6 Å corresponding to the bulk crystal structure of $Pr_2Ir_2O_7$[28]. We also observe topographs with atomically resolved terraces separated by ~ 6 Å, corresponding to the unit-cell along the [111]-direction, within which small islands (2.3 Å high) are seen (Fig.1d). The Fourier transform of both topographs show 3-fold symmetric Bragg peaks corresponding to the Kagome lattice structure (right inset of Fig.1c). The chemical nature of the surface termination in the topographic images is identified by carrying out DFT calculations of the surface stability and electronic density (see Supplementary Note 2). We find the lowest-energy cleaving plane to be midway between the metal layers because it cuts through only one of the four metal-oxygen bonds for each surface oxygen atom. The high calculated energy of alternative terminations suggests that the (111) surfaces have the natural 4:3 ratio of oxygen coverage with negligible defects on the respective Pr-rich and Ir-rich sides. DFT simulation of the STM surface termination further reveals an asymmetry in the surface electronic density between the Pr-Ir and Ir-Pr step heights that enables us to unambiguously identify terraces/islands in Fig.1d as Pr-rich/Ir-rich chemical composition. Regardless of the surface topology, however, the topographs show nanoscale inhomogeneity with picometer (pm) height variation (Fig.1c-g) corresponding to the variation in the integral local density of states (LDOS) in this stoichiometric material system. Spatial inhomogeneity is inherent to the complexity of correlated electron systems, such as manganites and cuprates, where various near-degenerate electronic phases can be continuously tuned at the nanoscale as a result of minute perturbations[29]. What causes this minute perturbations? The most likely candidate is intrinsic and antisite disorder, which is common in the pyrochlore iridate family with small ratio of $Ln^{3+}/Ir^{4+}$ ionic radii in $Ln_2Ir_2O_7$ as occurs in $Eu_2Ir_2O_7$[30]. However, for $Pr_2Ir_2O_7$, the ionic radius ratio $Pr^{3+}/Ir^{4+}$ = 1.8[31] is the largest in the family, indicating a minimal amount of antisite disorder. Yet, due to the unique band structure of $Pr_2Ir_2O_7$, which displays a node with quadratic bands touching at the Fermi energy[4], as further discussed below, minute perturbations can have dramatic impact on the LDOS.

The origin and nature of the spatial inhomogeneity is best visualized through spectroscopic imaging with the STM. Figure 2a-c show conductance (dI/dV) measurements on a 40 nm area at an energy of -10 meV below $E_F$ at different temperatures (and different fields of view).  Figure 2b, c corresponds to the same field of view of the topographs in Fig.1c, d, respectively. At all temperatures, a clear electronic inhomogeneity with nanometer length-scale, corresponding to relatively high and low density of states (at -10 meV) is observed. Quite remarkably, this nanoscale inhomogeneity is decoupled from the surface structural features (islands and step edges in Fig.1d) observed in the topograph suggesting the inhomogeneous puddles are extended to the bulk (see supplementary Note 3). For example, overlaying the unit-cell step and the Ir-rich islands seen in Fig.1d onto the conductance map of Fig.2c (gray lines) clearly reveals the absence of a link between the two (cross-correlation of -0.15), indicating the inhomogeneity not to be originating from the surface topographic structures and likely being electronic in nature. Moreover, the resemblance of the observed patterns in the conductance maps (Fig.2a-c) obtained from areas with very different topographic features (Fig.1c, d), further suggests that the electronic inhomogeneity is intrinsic to the bulk of the material.

To understand the origin of this electronic inhomogeneity, in Fig.2d-f we plot the dI/dV spectra, which corresponds to the LDOS, crossing two domains (black arrows in Fig.2a-c). At low temperature, the spectra reveal a sharp resonance right below the Fermi energy, originating from the orange region, that smoothly weakens and disappears before transitioning into a partial gap-like structure in the green region. Clearly, the overall spectral shape is very different in these two domains, indicating that the inhomogeneity corresponds to nanoscale phase separation of two different electronic states. As temperature is increased to 33 K, the spectra are broadened, yet maintain the same spatial behavior. By 45 K, thermal broadening takes over and only a hint of the resonance remains.

To obtain an unbiased and global description of the spatial phase separation, we use machine learning with K-means clustering algorithm that is based on the Euclidean separation of the data-space to spatially identify the different spectral clusters[32]. This method effectively sorts spectra whose features are most distinct (see Supplementary Note 4). Figure 3a-c shows the results of the K-means, revealing three clusters in all the data. The overall K-means images are analogous to the STM conductance maps at a bias of -10 meV (see Supplementary Note 5). This is not true for all energies, as the STM set-point effect dramatically alters the maps at energies between the set-point bias and the Fermi energy (see Supplementary Note 6 for a complete set of energy dependent conductance maps and their corresponding Fourier transforms). The K-means present the spatial phase separation originating from the entire spectral features (rather than a single energy) and independent of the experimental set-point effect (see Supplementary Note 4). Clearly, the spectra originating from the bright and dark domains represent two completely different electronic states with one showing a sharp resonance and overall higher density of states near $E_F$ as compared to the dark regions that show a partial gap. The average of the spectra originating from the different clusters are plotted in Fig.3d-f and compared to individual STM spectra from corresponding regions (Fig.3g-i). While the K-means analysis does result in three clusters, two of these clusters (green and black) are qualitatively similar in their spectral lineshape, yet quite distinct from that of the orange cluster.

The sharp asymmetric resonance revealed in Figs.2, 3 at low temperatures is reminiscent of the Fano lineshape that originates from Kondo hybridization and is the hallmark of heavy fermion Kondo-lattice material systems[33–37]. The nanoscale inhomogeneity of the Kondo hybridization in a stoichiometric material system, however, has not been observed in any of these Kondo lattice systems and may thus be a result of its competition with magnetic frustration on the Kagome lattice and its proximity to quantum criticality[2,23]. Note that the STM tunneling sensitivity to predominantly $f$-like and $spd$-like electronic states originating from tunneling into different exposed surfaces may also result in a peak and gap spectral lineshapes, respectively[38]. However, the appearance of the electronic phase separation on the very same atomically resolved surface, as well as the lack of correlation to the different exposed surfaces at all energies (see Supplementary Note 3), altogether excludes the tunneling sensitivity scenario as the origin of the observed phase separation. Furthermore, inhomogeneity due to intrinsic disorder, as seen for example in doped Kondo lattice systems[39–41], reveal the spectra to be only slightly modifies near dopants and their lineshapes remain qualitatively similar unlike the observed peak/gap lineshapes and their phase separation in the current study.

In $Pr_2Ir_2O_7$, Kondo physics is expected to originate from the Pr $4f$ electronic states through their hybridization with the Ir $5d$ itinerant electrons. To extract the Kondo temperature, we fit the data originating from the orange clusters to the Fano lineshape (see Fig.2 and Supplementary Note 7). The temperature-dependent resonance linewidths extracted from all the data yields a Kondo temperature with a Gaussian distribution centered at ~ 60±20 K (uncertainty corresponding to the standard deviation), in agreement with the resistivity minimum, providing spectroscopic evidence of inhomogeneous and frustrated Kondo-lattice in $Pr_2Ir_2O_7$. The break-down of the Kondo-singlet formation of the $f$-moments in the metallic state (green-black phase) together with the absence of static magnetic order in $Pr_2Ir_2O_7$ (as seen in neutron scattering and magnetization measurements)[3,21,42] therefore indicate that the $4f$ local moments in the non-Kondo phase (green-black phase) exhibit self-screening through long-range entanglement among themselves. Whether these domainsare responsible for the anticipated MSL state that emerges below T = 1.7 K requires lower-temperature measurements. In these domains though, the LDOS displays a square-root energy dependence pinned at the Fermi energy (as seen by the partial gap in the dark-green domains) originating from the Fermi node and the quadratic band touching[4]. The location of Fermi energy exactly at the node supports the charge neutrality and therefore the stoichiometry of our samples (see Supplementary Note 8).

Spontaneous emergence of multiphase electronic states, near critical points, forming nanoscale domains in an otherwise stoichiometric quantum material system is the hallmark of complexity in correlated electron systems where tiny perturbations of spin, charge, lattice, or orbital degrees of freedom can have a large response on the electronic ground state[29,43]. Near continuous phase transitions (see Supplementary Note 9), the spatial organization of the electronic states often follows scale-invariant fractal domain structure, where various physical quantities display power-law scaling over multiple decades[44,45]. Following very recent work on the AFM $NdNiO_3$[46] as well as earlier work on cuprates[45,47,48], we study the pattern formation through the spatial organization of the electronic domains. To obtain a meaningful analysis of the geometric clusters and their critical exponents, we first carry out a high resolution spectroscopic imaging on the same field of view of Fig.1c (see Supplementary Note 10) that corresponds to an area of 65 nm (at T = 33 K) with a single pixel resolution of 0.25 nm (see Supplementary Note 10). Similar analysis at T = 10 K, yet with lower spatial resolution, is also presented in Supplementary Note 10.

Statistical analysis of domain geometry using the STM conductance maps at some energy, however, suffers from the experimental set-point effect which renders the integral of the conductance between the set-point bias and the Fermi energy to be spatially uniform. To overcome this technical issue, we use the K-means map which clusters the spectra based on their mathematical resemblance and independent of the set-point effect. To binarize the K-means map, the machine learning algorithm is run by a default two clusters (Fig.4a, which corresponds to the conductance map in Supplementary Fig.S17b). We use the box-counting method as well as the geometric domain distribution analysis (see Supplementary Note 11) to analyze the spatial patterns. Figure 4b displays the distribution of number of boxes (N) versus (box size)$^{-1}$ in a log-log plot, which is a measure of the fractal dimension. Here, N corresponds to the number of boxes that cover the domains. Over two-and-a-half decades, the distribution is linear and reveals a scale-invariant power law $N \propto (\frac{1}{boxsize})^D$ with a fractal dimension D = 1.49 ± 0.02 (Fig.4c). The fractal dimension is also obtained by analyzing the domains' area (A), perimeter (P), and area distribution (D(A)), where the perimeter scales as $P \propto A^{D/2}$, and yields D = 1.53 ± 0.03 (Fig.4c) in agreement with the box counting method. On the other hand, the domains' area distribution scales as D(A) $\propto A^{-\tau}$ as the geometric clusters become near-critical. Fig.4e shows the power-law behavior of the distribution that persists for over two decades of scaling with the Fisher exponent $\tau$=1.62 ± 0.16 (Fig.4d). We further examine the scaling behavior by extracting the radius of gyration (R) of the domains, which is expected to display power-law behavior as a function of the perimeter (P) and area (A) near critical points[44,45,47]. From Figure 4e,f, the power-law behavior is evident revealing the hull and volume critical exponents $d_h$ = 1.39 ± 0.03 and $d_v$ = 1.76 ± 0.06. Finally, following the earlier work[45–48], we calculate the pair-connectivity function for the map in Fig.4a for both clusters (see Supplementary Note 12). The pair-connectivity is a measure of the probability that two different sites separated by a distance $r$ on the map belong to the same cluster. Fig.4g shows the extracted data and fits to power-law with exponential cut-off $r^{-\eta}e^{-r/\chi}$, where $\eta$ and $\chi$ are the scaling exponent and the correlation length, respectively. The pair connectivity function displays a power law behavior for the black clusters (MSL; $\eta$ = 0.06) with correlation length of 15 nm in contrast to the orange cluster (HFL) which dominates the map with a correlation length of over 1000 nm (larger than the map itself) indicating that the system is in proximity to a critical transition but is located on the HFL side of the phase diagram.

The power-law scaling over more than two and a half decades of the various physical quantities displayed in Fig.4 indicates a scale-invariant fractal geometry in proximity to a critical point. The values of the different extracted exponents, however, cannot be explained by the uncorrelated percolation theory indicating that strong electronic correlations originating from magnetic frustration are in play in the formation of the spatial geometric patterns.

Our work reveals a remarkable new route for tuning quantum phases at the nanoscale. It also raises the question of what microscopic physics allows such an exquisite tuning. We suggest that the sensitivity of this tuning is amplified by the dilute carrier nature of $Pr_2Ir_2O_7$. ARPES shows that the Fermi energy of the conduction electrons lies close to a quadratic band-touching[4]. Using this band structure as an input, we show in Fig.4h that relatively small variation of the electron filling ($\mu/D$, where $\mu$ is the chemical potential and $D$ is the bandwidth), caused by the disorder potential, can produce a sizable change to the bare Kondo scale $T_K^0(\mu)$. This variation is amplified by the closeness of Fermi energy to the quadratic band-touching, in contrast to the generic case, where the Fermi energy is deep in the middle of a conduction-electron band (see Supplementary Note 13), which highlights the amplified ability of using the disorder potential to tune the ratio between the Kondo and RKKY energy scales.

In conclusion, the findings in this work reveal the rich and complex behavior of emergent electronic states and their near-critical nature in a rare quantum material where magnetic frustration and Kondo quenching are considered on an equal footing. Moreover, it is important to state that the existence of interweaved domain boundaries between the different electronic phases discovered in this work and the domain-wall states along these boundaries may play a relevant role in the emergence of the chiral spin-liquid and the anomalous Hall effect observed at much lower temperatures[5]. Future experiments at sub-Kelvin temperatures, which are beyond the scope of this initial study, may elucidate the role of the electronic phase separation in the emergence of the AHE as well as the proximity of $Pr_2Ir_2O_7$ to the delocalization-localization transition line (orange line in Fig.1a) in the global phase diagram.

## Methods

### Sample Preparation

Single crystals of $Pr_2Ir_2O_7$ of size approximately 1 cubic millimeter were grown using a flux method. Samples of roughly $1 \times 1 \times 1$ $mm^3$ were attached to metallic plates using H74F epoxy. A conducting channel made of H20E silver conductive epoxy was formed from the plate to the side of a sample. An aluminum post was attached to the top surface of the sample perpendicular to the (111) cleaving plane using H74F epoxy.

### Scanning Tunneling Microscopy

Samples are cleaved in situ in ultra-high vacuum at room temperature by knocking off the aluminum post. It is then immediately transferred to the STM where it is placed next to a Cu(111) crystal which is used to prep the PtIr tips before each experiment. They are both cooled down to the desired experimental temperature. The Cu is treated prior to tip preparation by several rounds of sputtering and annealing. STM topographs are taken in constant current mode, and the dI/dV measurements are performed using a lock-in amplifier with a reference frequency of 0.921kHz.

### Theoretical modeling

We describe the calculation to study the effect of tuning the local potential on the Kondo scale in the case of quadratic touching conduction-electron bands. Two conduction-electron bands ($c_1$ and $c_2$), each with a quadratic dispersion, touch at energy 0 (Fig. 4h, inset). The local potential is modeled in terms of an effective chemical potential, $\mu$. The bands interact with a spin-1/2 local moment by the Kondo coupling $J_K$. The model is solved in terms of saddle point equations that are exact in the limit of large N, where N corresponds to generalizing the SU(2) spin symmetry to SU(N). In the calculation, N is set to the physical value 2. We use the pseudo-fermion representation, in which the local moment is described by $\boldsymbol{S} = \frac{1}{2}f^\dagger \boldsymbol{\sigma} f$, where $\boldsymbol{\sigma}$ are the generators of the SU(N) group, along with the constraint by $\sum_\sigma f^\dagger f = N/2$. A Hubbard Stratonovich transformation is used to decouple the Kondo coupling term, which yields an effective hybridization: $\sum_{\sigma,p} f_\sigma^\dagger \left( B_1 c_{1,\sigma}(p) + B_2 c_{2,\sigma}(p) \right) + h.c.$. Here, $B_1$, $B_2$ are the auxiliary fields for the decoupling, and become complex numbers in the large N limit. We minimize the ground state energy of the effective model, subject to the pseudo-fermion constraint. The conduction-electron bands correspond to a pseudogap in the total density of states, with $\rho(E) \propto |E/D|^{1/2}$. We set the half bandwidth D=1; for each band, we normalize $\rho(E)$ so that its area is 1. For illustrative purpose, we focus on the Kondo coupling $J_K = 0.5D$. The Kondo energy scale is derived by calculating the shift of the energy due to the Kondo coupling: $E_K^0 \equiv k_B T_K^0 = E(J_K) - E(J_K = 0)$.

## Data availability

The authors declare that the main data supporting the findings of this study are available within the article and its Supplementary Information files. Extra data are available from the corresponding author upon request.

## Acknowledgements

We acknowledge discussions with Erica Carlson and Karin Dahmen.

P.A. acknowledges support from the U.S. National Science Foundation (NSF) CAREER under Award No. DMR-1654482. S.H. and A.N.K. acknowledge the NSF support (Award No. DMR-1821815) and the Extreme Science and Engineering Discovery Environment computational resources (NSF award No. ACI-1548562, project No. TG-PHY190024). The work at Rice University was in part supported by the NSF (DMR-1920740) and the Robert A. Welch Foundation (C-1411). This work is partially supported by CREST (JPMJCR18T3), by Grants-in-Aids for Scientific Research on Innovative Areas (15H05882 and 15H05883) from the Ministry of Education, Culture, Sports, Science, and Technology of Japan, and by Grants-in-Aid for Scientific Research (16H02209, 16H06345, 19H00650) from the Japanese Society for the Promotion of Science (JSPS).

## Author Contributions

M.K. and J.F. performed the STM measurements. M.K., J.F. and M.G. performed the data analysis. Q.Y. and S.N. synthesized and characterized the materials. K.S. and M.L. performed the K-means analysis. S.H. and A.N.K. performed the DFT calculations. H.H., S.E.G., and Q.S. performed the theoretical calculations. P.A. designed the project and wrote the manuscript. All authors commented on the manuscript.

## Competing Interests

The authors declare that they have no competing interests.

**Figures**

Figure 1

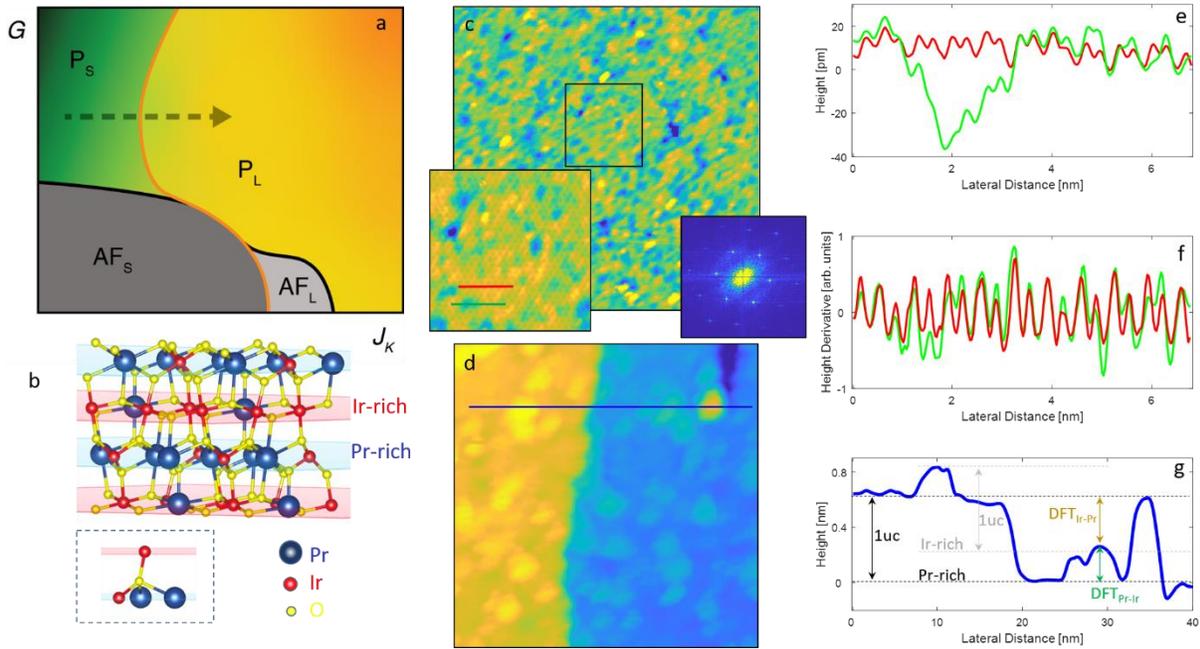

Fig 1. (a) Global phase diagram of frustrated Kondo lattice. $P_S/P_L$: Non-magnetic phase with small/large Fermi surface, $AF_S/AF_L$: Antiferromagnetic phase with small/large Fermi surface. (b) Crystal structure of $Pr_2Ir_2O_7$ along the (111) direction highlighting the Pr-rich (light blue plane) and Ir-rich (light red plane) surfaces. The bottom image shows the four oxygen tetrahedron between the Ir-rich and Pr-rich planes with three interlayer and one intralayer oxygen-metal bonds. Based on our DFT calculations, cleaving breaks the single intralayer bond keeping the oxygen atom intact to its layer. (c) Topographic image (65 nm, temperature of 33 K, set-point bias of 100 mV, set-point current of 500 pA) of the cleaved (111) surface showing an atomically resolved lattice structure. The left inset is a zoom-in of the black box in the main image, showing more clearly the atomic structure. The right inset is the Fourier transform showing 3-fold symmetric Bragg peaks. (d) Topograph (40nm, temperature of 45 K, set-point bias of 100 mV, set-point current of 500 pA) showing atomically resolved terraces with a step height corresponding to the unit cell in the [111] direction, with small islands observable on the terraces. (e) Linecuts of the topograph along the red and green lines in the left inset of (c). (f) Height derivative of the linecuts in (e) indicating no missing atoms. (g) Linecut along the blue line in the topograph in (d), showing the main terraces formed by a Pr-rich surface and the islands formed by the Ir-rich surface. The dashed horizontal lines represent the DFT step height from a Pr-rich to Ir-rich terraces.

Figure 2

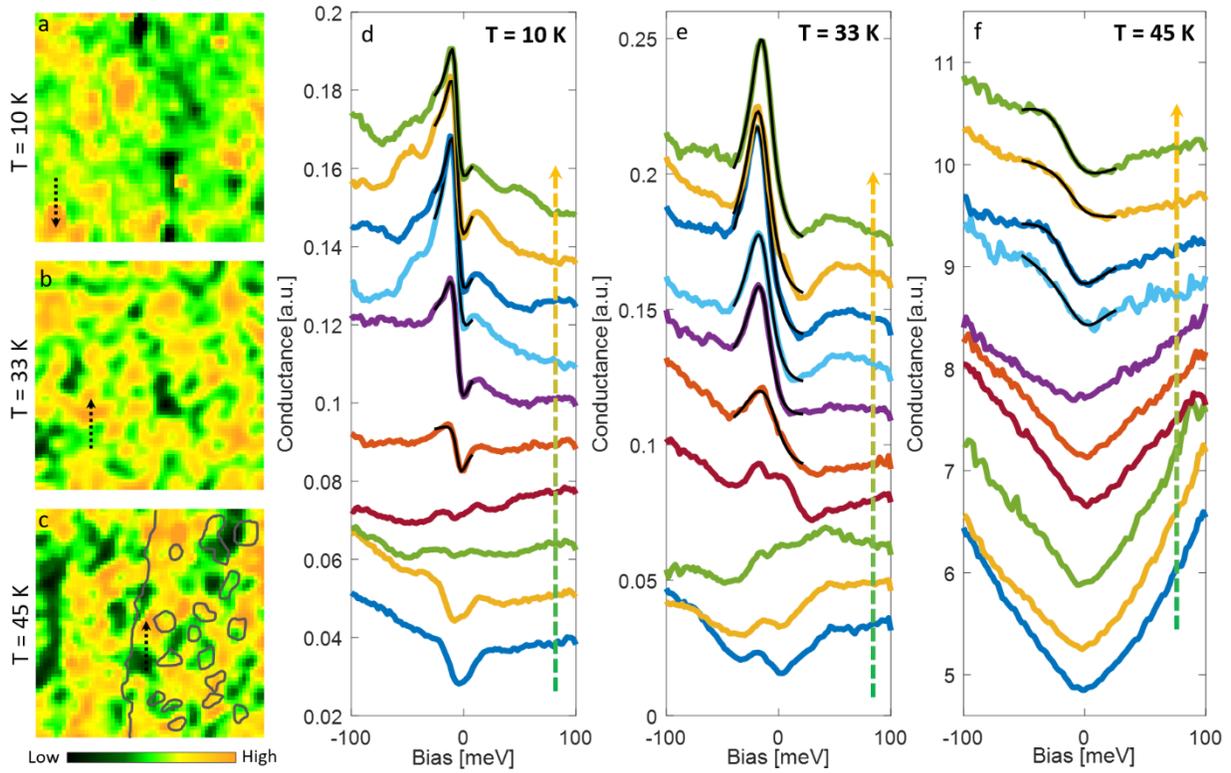

Fig 2. (a)-(c) STM conductance maps at three different temperatures on different fields of view. Each map is 40 nm in size and corresponds to the local DOS at an energy of -10 meV below $E_F$. set-point bias is 100meV for all maps. Set-point current is 200 pA in (a) and 500 pA in (b, c). A clear electronic inhomogeneity with nanometer length scale is observed for all temperatures. The gray lines in (c) are an overlay of the unit cell step and islands from Figure 1(d) demonstrating no link between the electronic and structural features. (d)-(f) Spectra taken along the black dashed arrows shown in the conductance maps. Each arrow crosses from one domain into the other. The arrows alongside the spectra mirror the directions of the black arrows.

Figure 3

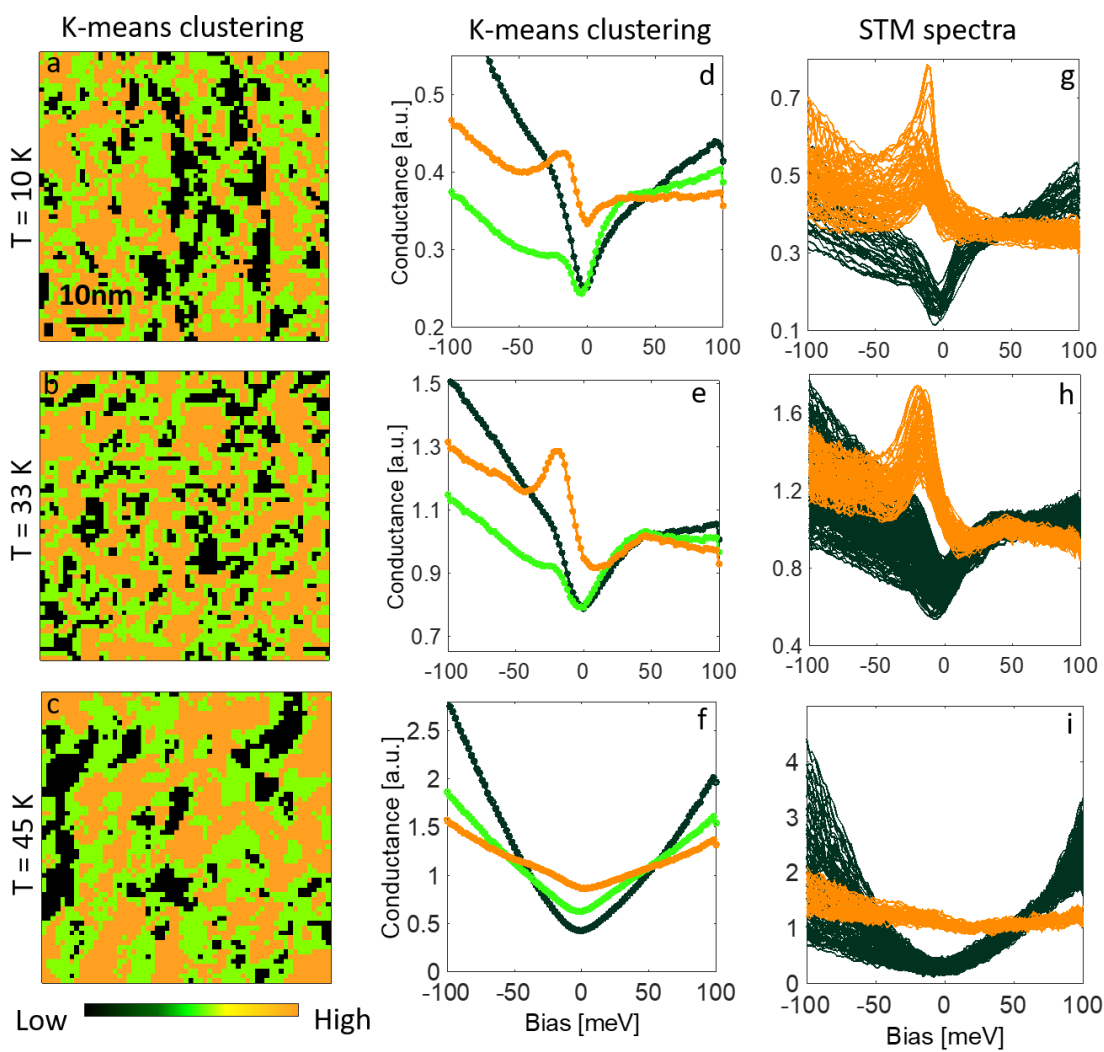

Fig 3. (a)-(c) K-means cluster maps for three temperatures corresponding to the maps shown in Fig.2(a-c). Each map exhibits three main clusters. All images are 40nm in size. (d)-(f) The average of all spectra in each of the clusters from the k-means map. (g)-(i) Individual spectra plotted from the orange and black areas of the conductance maps corresponding to the k-means maps. There is a clear difference between the spectra from these different areas at all three temperatures.

Figure 4

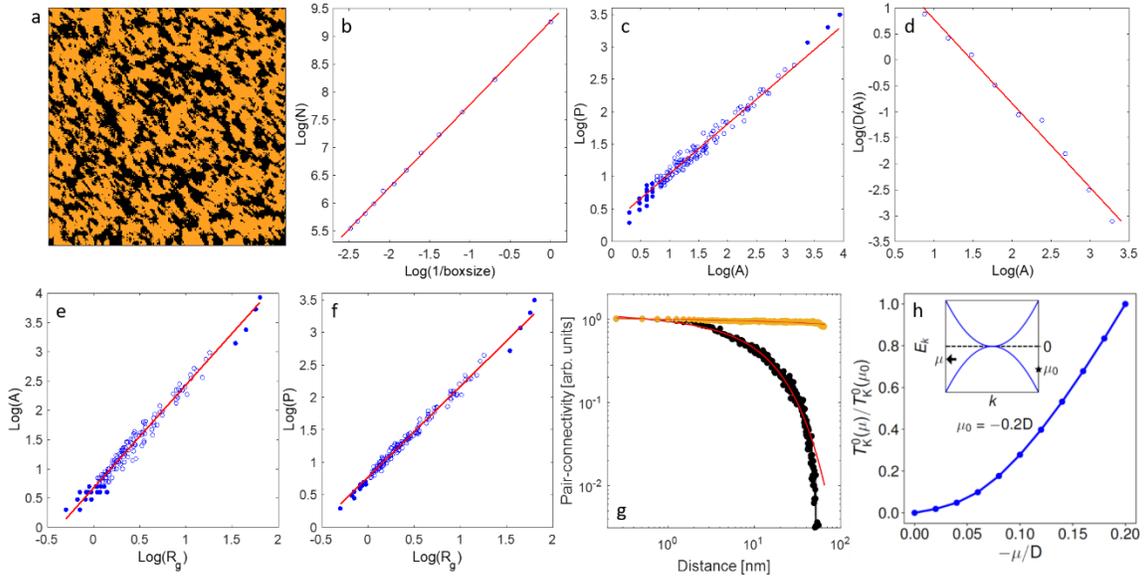

FIG. 4. (a) K-means cluster map (65 nm) forced using two clusters. The black clusters were used for the fractal analysis. The corresponding STM conductance map can be found in Supplementary Note 11. The data is taken at 33 K. (b) The fractal dimension analysis using a box-counting method resulting in fractal dimension D = 1.49 ± 0.02. The data points are fitted by the least square method to the power law $N = \left(\frac{1}{boxsize}\right)^D$ where $N$ denotes the number of boxes of clusters. (c) Perimeter vs area of the clusters. Filled circles are excluded from the fit. Using P = $A^{D/2}$ gives a fractal dimension value of D = 1.53 ± 0.03; this corresponds well with the box counting method. (d) Cluster area distribution with logarithmic binning, giving a Fisher exponent $\tau$ = 1.62 ± 0.16. (e) Utilizing the radius of gyration for each cluster, $R_g = \sqrt{\langle(r - \langle r \rangle)^2\rangle}$, where $r$ is summed over all points in that cluster, the equation $A = R_g{}^{d_v}$ gives a value of $d_v$ = 1.76 ± 0.06. (f) Similarly using P = $R_g{}^{d_h}$ gives a value of $d_h$ = 1.39 ± 0.03. (g) Pair connectivity function extracted from the black (black) and orange (orange) regions. The lines are fit to the function y = $r^{-\eta}e^{-r/\chi}$, where $\eta$ and $\chi$ are the scaling exponent and the correlation length. (h) Kondo scale $T_K^0(\mu)$ normalized by $T_K^0(\mu_0 = -0.2D)$, as a function of chemical potential $\mu$ in case of a quadratic touching conduction electron bands. The Kondo scale is seen to be highly sensitive to slight variations of μ near the node (Fermi energy) in the band structure. The inset illustrates the two conduction-electron bands, each with a quadratic dispersion, touching at the Fermi energy.

15